\documentclass[conference]{IEEEtran}

\usepackage{url}
\usepackage{csquotes}
\usepackage{xprintlen}
\usepackage{todonotes}
\usepackage[official]{eurosym}
\usepackage{enumitem}
\usepackage{tabularx}
\usepackage{amssymb}
\usepackage{siunitx}
\usepackage{tabularx}
\usepackage{csquotes}
\usepackage{float}
\usepackage{listings}
\usepackage[skip=10pt, font=small]{caption}
\usepackage{tikz,lipsum}
\usepackage[many]{tcolorbox}  

\definecolor{main}{HTML}{5989cf}
\definecolor{sub}{HTML}{eeeeee}
\newtcolorbox{boxC}{
    left=3pt, 
    right=3pt,
    breakable,
    colback = sub,
    boxrule = 0pt 
}

\begin{document}

\title{Résumé-Driven Development: A Definition and Empirical Characterization}

\author{
    \IEEEauthorblockN{
        Jonas Fritzsch,
        Marvin Wyrich,
        Justus Bogner,
		Stefan Wagner
	}
	\IEEEauthorblockA{
        University of Stuttgart, Germany, Institute of Software Engineering\\
    	\{firstname.lastname\}@iste.uni-stuttgart.de
    }
}

\maketitle

\begin{abstract}
Technologies play an important role in the hiring process for software professionals.
Within this process, several studies revealed misconceptions and bad practices which lead to suboptimal recruitment experiences.
In the same context, grey literature anecdotally coined the term \textit{Résumé-Driven Development} (RDD), a phenomenon describing the overemphasis of trending technologies in both job offerings and resumes as an interaction between employers and applicants.
While RDD has been sporadically mentioned in books and online discussions, there are so far no scientific studies on the topic, despite its potential negative consequences.
We therefore empirically investigated this phenomenon by surveying 591 software professionals in both hiring (130) and technical (558) roles and identified RDD facets in substantial parts of our sample: 60\% of our hiring professionals agreed that trends influence their job offerings, while 82\% of our software professionals believed that using trending technologies in their daily work makes them more attractive for prospective employers.
Grounded in the survey results, we conceptualize a theory to frame and explain Résumé-Driven Development.
Finally, we discuss influencing factors and consequences and propose a definition of the term.
Our contribution provides a foundation for future research and raises awareness for a potentially systemic trend that may broadly affect the software industry.
\end{abstract}


\IEEEpeerreviewmaketitle

\section{Introduction}

\noindent
\textit{\enquote{Specific technologies over working solutions, hiring buzzwords over proven track records, creative job titles over technical experience, and reacting to trends over more pragmatic options}}~\cite{rdd.io2020} \textendash \hspace{3pt}this is the future of software development if we would adhere to this satirical manifesto for \textit{Résumé-Driven Development}.
According to its signatories, the phenomenon is already present today.
This invites the question if there is more than anecdotal evidence to it.

Ebert and Counsell state in their projection of the state of Software Technology in the year 2050 that \textit{\enquote{Software-driven systems are exhibiting rapidly growing complexity [...] based on a fast-changing technology landscape}}~\cite{Ebert2017}.
They further project that \textit{\enquote{overwhelming complexity, combined with insufficient development competences, greatly decreases software-driven products’ quality}} which would become evident in e.g. \textit{\enquote{public’s decreasing acceptance of [...] self-driving cars}}~\cite{Ebert2017}.
Already a decade ago, Adomavicius et al. regarded the sheer number of available technologies and the complex set of relationships among them as a major challenge~\cite{Adomavicius2008}.

The recent Stack Overflow 2020 Developer Survey among 65.000 software developers found that \textit{\enquote{around 75\% of respondents [...] learn a new technology at least every few months or once a year}}~\cite{Stackoverflow2020}. 
The survey also revealed that technologies are regarded as the most important job factor by software professionals.
This focus may come at the expense of other skills, leading to \textit{\enquote{insufficient development competences}} \cite{Ebert2017} as Ebert and Counsell state it.

In the context of the software professional recruiting process, lively discussions have come up in developer forums, blogs, or social media where occasionally the terms \textit{Résumé-Driven Development} (RDD)~\cite{HackerNews2017a, Loukides2014, Laptev2019} and similarly \textit{CV-Driven Development}~\cite{Teller2020} were used.
Classon associates this notion with selecting \textit{\enquote{tech stacks, architecture, methodologies, and protocols based on what looks good on the resume}}~\cite{Classon2019}, while Ford et al. describe it as a pitfall by architects becoming enamored in latest technologies~\cite{Ford2017a}:
\textit{\enquote{utilizing every framework and library possible to tout that knowledge on a resume}}.
The topic tends to provoke heated and even polemic debates, as e.g. visible in~\cite{HackerNews2017} or~\cite{Jha2020}.

While the term RDD has been sporadically used in books and online discussions, we have not found any empirical investigation of the phenomenon nor a definition and theory of the term in scientific literature.
Thus, we see a need for further investigation and clarification. Software development professionals and decision makers will equally benefit from a more in-depth understanding of a phenomenon that may be systemic in nature and has the potential to substantially affect software development practice.
In this study, we identify stakeholders, their intentions and relationships, as well as other influencing factors constituting to the phenomenon \textit{Résumé-Driven Development}.
As a result, we provide a definition and theoretical construct to substantiate this notion.
Consequently, we formulate our research objective as follows:

\medskip

\noindent
\textit{Conceptualize a theory for Résumé-Driven Development\\
For the purpose of framing and explaining the phenomenon\\
With respect to its scope, influences, and potential impact\\
From the viewpoint of human resource, software professionals, and computer science students\\
In the context of the software industry}

\medskip

\noindent
To refine our objective and to set the scope for a more detailed investigation, we formulate several research questions.
In this context, we distinguish between a hiring and an applicant perspective.
While software professionals sometimes act in both roles, hiring professionals and computer science students are commonly associated exclusively with one perspective.

\medskip
\begin{enumerate}[label=RQ\arabic*:,leftmargin=*]
    \item How valuable does hiring perceive technology-related characteristics of an applicant's skill set, i.e. \enquote{breadth vs. depth} and \enquote{established vs. latest}?

    \item To what degree is hiring influenced by latest / trending technologies when creating job offerings?

    \item To what degree are applicants influenced by latest / trending technologies when refining their resumes?
    
    \item What experiences did applicants have in practice when using latest / trending technologies? 
\end{enumerate}
\medskip

\noindent
To answer these research questions, we designed and conducted an exploratory survey among software professionals with both hiring and technical roles as well as computer science students.
A total of 591 participants could be recruited.
Based on the statistical analysis of the survey results, we developed a theory for analyzing and describing \textit{Résumé-Driven Development} (RDD) following the framework for theories in information systems proposed by Gregor \cite{Gregor2006}.

\section{Background and Related Work}
\label{sec:background}
\noindent
In this section, we define the problem area and present existing anecdotal evidence, which leads to a vague understanding of RDD.
We characterize the involved stakeholders and put our research into the context of similar phenomena and related work.

\subsection{Problem Area and Scope}

\noindent
We first define stakeholders and their relationships as objects of investigation.
As an abstraction, we declare stakeholders representing a company or employer as the \textit{hiring} perspective.
Usually, this responsibility lies with people managers, team leads, hiring professionals, or head hunters in the software industry.
On the opposite side, we define software professionals (developers and related job roles) as employees in the \textit{applicant} perspective.
Computer science students are also covered by this perspective.
For the relationship between both perspectives within the recruiting process, the following artifacts are of particular importance: job advertisements issued by \textit{hiring} and resumes or curricula vitae (CVs) submitted by \textit{applicants}.
The desired / promoted skills with regards to technologies play an important role for job advertisements and resumes in software development~\cite{Stackoverflow2020}. 

Ideally, the technologies in a job advertisement reflect the company's actual demands.
Likewise, we would expect applicants to promote their actual skills and experiences in their CVs.
Potential problems arise if expectations or assumptions of the other perspective consciously or unconsciously influence the creation of these documents.
Improperly prioritized or advertised information could lead to misunderstandings in the short term and more severe consequences in the long term.

To investigate how technology selection is influenced by these forces, we distinguish between \textit{established} technologies and \textit{latest / trending} technologies.
Instances of the latter (also referred to as \textit{hyped} technologies) can be frameworks, programming languages, middleware, concepts, or architectural styles which have recently gained excessive industry attention, e.g. through print / online media, conferences, social media, video platforms, or blogs.
Some concrete examples are microservices, IoT, DevOps, Docker, Serverless, Apache Kafka, Go, Kotlin, Kubernetes, Angular, Vue.js, TensorFlow, or Blockchain technologies.
In contrast, established technologies are of a more conservative nature, widely accepted since years, and a large body of knowledge exists on how to use them correctly.

\subsection{Related Phenomena and Studies}

\noindent
We distinguish the meaning behind \textit{Résumé-Driven Development} from the phenomenon of \textit{Hype-Driven Development}~\cite{Kirejczyk2016,HackerNews2017b,Redd2018}, which refers to a similar concept but on a more general level.
There, trending technology is not seen as an influence to the job market, but for alleged product optimizations, e.g. to proactively plan ahead.
Another related phenomenon is \textit{Cargo Cult Software Engineering}~\cite{McConnell2000}.
It is characterized by the blind use of code, design patterns, or technologies without fully understanding the reasons behind it.
Here, the motivation to adopt a technology or practice is based on having seen its successful application somewhere else.

In recent years, several studies have revealed flaws in the hiring process of tech companies.
These studies examine how the skills of potential candidates are tested and which biases are involved.
In particular, such studies address wrong assumptions about what interviewers expect~\cite{Ford2017a}, psychological aspects of solving coding challenges~\cite{Behroozi2018, Behroozi2020a}, or the impact of individual characteristics in this context~\cite{Wyrich2019a}.
The results cast reasonable doubts on the ability of current interview practices to fairly find suitable candidates for building diverse software teams~\cite{Behroozi2020}.

The recent Stack Overflow Developer Survey 2020 showed that \textit{languages, frameworks, and other technologies}~\cite{Stackoverflow2020} are by far the most important job factor for developers.
According to Gant~\cite{Gant2019}, developers tend to \textit{\enquote{fixate on programming languages and technology stacks as measurements of their value}}, although it would be possible to \textit{\enquote{build a perfectly acceptable application using tools that aren’t popular}}.
More than one-third of the 65,000 Stack Overflow's survey participants learn a new technology every few months and another third at least once a year.
A complementary study by Montandon et al.~\cite{Montandon2020} on more than 20,000 Stack Overflow job postings also confirmed the importance of this aspect for the hiring perspective.
They found expertise in specific programming languages and frameworks to be the most required hard skill by companies.

While existing studies show that developers frequently learn new technologies and that specific technologies are one of the most frequent requests in job advertisements, a systematic study that links these concepts and examines their interaction is still missing.
We do not yet know in which ways companies and software developers are influenced by technology trends and if developers learn new technologies to stay competitive or mostly out of pleasure and self-motivation.
With our study, we aim to fill these gaps and to shed light on the influence of latest and trending technologies from both the hiring and the applicant perspective.

\section{Research Design}

\noindent 
In this study, we conducted basic research which aims to provide understanding rather than a solution to the problem ~\cite{Wohlin2015a}.
We applied inductive logic, also referred to as a bottom-up approach to theory-building~\cite{Bhattacherjee2012}.
Starting from related literature, we collected background information and insights about the problem area.
After identifying stakeholders and relationships among them, we designed and conducted an exploratory survey to gain insights for both perspectives.
We collected quantitative data via a questionnaire and used statistical methods for the analysis.
To ensure scientific rigor, we follow the seven-step guideline for survey design by Kasunic~\cite{Kasunic2005}, complemented by Linåker et al.'s more recent annotations \cite{Linaker2015}.
Finally, we derived general conclusions resulting in a theory about the phenomenon RDD.

\subsection{Sampling Strategy}

\noindent 
Our target population is determined by the above defined perspectives, namely practitioners and students in a software development context with hiring or technical roles.
For distributing the survey, we relied on convenience (or accidental) sampling which is a common approach in software engineering~\cite{Linaker2015}.
Because such a sample may not be representative, we used several distribution channels in parallel to reach a larger and more diverse audience~\cite{Ciolkowski2003}.
We distributed the survey via a leading German online magazine for software developers (Heise), professional platforms (LinkedIn, XING), social networks (Twitter), and developer forums (Reddit).
Moreover, we contacted about 100 IT and software companies which participate in a recurring career fair at our university.
Using mailing lists, we also reached out to enrolled students of several computer science courses of our and a neighboring university.
Lastly, personal contacts of all involved researchers were invited to participate.
Participants were also encouraged to forward the survey to their professional contacts in the software industry.

\subsection{Questionnaire Design}
\label{QuestionnaireDesign}

\noindent 
To cover both perspectives, we designed a twofold web-based survey.
Participants were able to choose their perspective (\textit{hiring}, \textit{applicant}, or both) at the beginning.
In the survey, we used the term \textit{technical} for the \textit{applicant} perspective to make the selection clearer.
For creating the questionnaire, we transformed each of the research questions into several survey questions.
We used mostly closed-ended descriptive questions to ease analyzability.
Because we were interested in attitudes and opinions, the majority of questions relied on a 5-point Likert scale~\cite{Preedy2010} to measure agreement (items ranged from \enquote{strongly disagree} to \enquote{strongly agree}).
Additional demographic questions referred to aspects like job role, years of professional experience, or company domain.
All non-free-text questions were mandatory.
However, for sensitive questions, we provided a \enquote{Prefer not to answer} option.
Answering the survey took approximately 5 minutes.
The complete questionnaire was in English and we used the EvaSys SurveyGrid application~\cite{ElectricPaperEvaluationssystemeGmbH2020} to create and host it.
% Before entering the survey, each participant needed to give consent for publishing the collected data in an aggregated form.
We ensured anonymity to all participants and consequently removed all personal information from the published data set.

\subsection{Pilot Test and Data Collection}

\noindent 
Four members of our research group reviewed the questionnaire, which led to several refinements in wording and layout.
To test the analysis procedure, the questionnaire was populated with random data in a simulation run.
The results were then used to pilot the analysis script, which again led to improvements for both the script and the questionnaire.
The final survey was activated on May 14th, 2020 with two months of data collection until July 14th, 2020.
To minimize the risk of outages of the hosting service, we sent out the invitations staggered over several days.
We regularly monitored response rates and reminded personal contacts after two to four weeks.
As an additional incentive, we pledged to donate \euro{1.00} to UNICEF for each of the first 100 responses, thereby supporting children's rights to an education.

\subsection{Statistical Analysis and Theory Building}
\noindent
After retrieving the results, the first step was data cleaning.
We removed incomplete responses, incomplete answers (e.g. for the ranking questions), unnecessary columns, and personal information from the final comment field. As a result, we obtained 591 valid responses out of 593.
In addition to data type and missing value harmonization, we also resolved the free-text answers into consistent values (e.g. country or \enquote{other} options).
With the final data set, we then first used descriptive statistics and Likert plots to analyze the distributions.
For inferential statistics, we started with an exploratory analysis via a correlation matrix (Spearman).
These results were then used to build and continuously refine linear regression models.
Lastly, the models were successfully tested for required properties like the absence of heteroskedasticity (Breusch-Pagan test with $p >> 0.05$).

For our theory, we relied on the taxonomy proposed by Gregor~\cite{Gregor2006}.
Her work provides a seminal framework for theory building.
In our study, we conceptualized a \textit{theory for analyzing}, also called \textit{Type I} theory.
Such theories \textit{\enquote{describe or classify specific dimensions or characteristics of individual groups, situations, or events by summarizing the commonalities found in discrete observations.
They state \enquote{what is}}}~\cite{Gregor2006}.
Descriptive theories are needed when nothing or very little is known about the phenomenon in question.
Compared to higher theory types, they do not emphasize causal explanations.
Due to the unexplored nature of RDD, we focus on the description of observed characteristics and relationships between them.
Future studies will be needed to support our theory.

\section{Survey Results}

\noindent 
A wide range of hiring and software professionals participated in our survey.
In total, we received 593 responses of which 591 were valid.
Due to the various distribution channels, we can only provide an estimate for the response rate based on our regular monitoring.
For contacts approached via personal email invitations or distribution lists, we saw timely responses for 5 to 10\% of the addressees.
Of the 591 valid responses, 130 answered for the \textit{hiring} and 558 for the \textit{applicant} perspective, i.e. 97 participants answered for both perspectives.

\begin{table}[!ht]
    \centering
	\caption{Responses per Perspective and Role}
	\label{table:dems_perspective_role}
	\begin{tabular}{p{0.25\textwidth}rr}
		Role & \# Hiring & \# Applicant\\ 
        \hline
        \hline
        Software Engineer & & 193\\
        People Manager / Team Lead & 115 & 27\\
        CS Student & & 102\\
		Architect & & 76\\
        Consultant & & 36\\
        Researcher / Lecturer & & 33\\
        System Administrator & & 29\\
        DevOps Engineer & & 27\\
        Project Manager & & 20\\
        Head Hunter & 8 &\\
        Test / Quality Engineer & & 8\\
        Requirements Engineer & & 7\\
        HR Professional & 7 &\\
        \hline
        Overall & 130 & 558\\
		\hline
	\end{tabular}
\end{table}

\begin{table}[!ht]
    \centering
	\caption{Professional Experience of Respondents}
	\label{table:dems_prof_experience}
	\begin{tabular}{
    	p{0.32\textwidth} 
    	>{\raggedleft}p{0.04\textwidth} 
        @{\hspace{1.5\tabcolsep}}
    	>{\raggedleft}p{0.05\textwidth} 
	}
		\# Years of Experience & \multicolumn{2}{r}{\# Respondents}\\ 
        \hline
        \hline
        None & 26 & (4.4\%) \tabularnewline
        2 or less & 93 & (15.7\%) \tabularnewline
		3-5 & 131 & (22.2\%) \tabularnewline
        6-10 & 103 & (17.4\%) \tabularnewline
        10-20 & 135 & (22.8\%) \tabularnewline
        21 or more & 103 & (17.4\%) \tabularnewline
        \hline
        Overall & 591 & (100\%) \tabularnewline
		\hline
	\end{tabular}
\end{table}

\begin{table}[!ht]
    \centering
	\caption{Industry Affiliation of Respondents}
	\label{table:dems_company_domain}
	\begin{tabular}{
    	p{0.32\textwidth} 
    	>{\raggedleft}p{0.04\textwidth} 
        @{\hspace{1.5\tabcolsep}}
    	>{\raggedleft}p{0.05\textwidth} 
	}
		Company Domain & \multicolumn{2}{r}{\# Respondents}\\
        \hline
        \hline
        Software \& IT Services & 246 & (41.6\%) \tabularnewline
		Automotive & 84 & (14.2\%) \tabularnewline
        Manufacturing \& Defense & 38 & (6.4\%) \tabularnewline
        Research \& Education & 37 & (6.3\%) \tabularnewline
        Finance \& Insurance & 31 & (5.2\%) \tabularnewline
        Public Sector & 21 & (3.6\%) \tabularnewline
        Healthcare & 19 & (3.2\%) \tabularnewline
        Retail & 13 & (2.2\%) \tabularnewline
        Hardware \& Semiconductors & 12 & (2.0\%) \tabularnewline
        Logistics \& Public Transport & 12 & (2.0\%) \tabularnewline
        Telecommunications & 10 & (1.7\%) \tabularnewline
        Energy \& Utilities & 9 & (1.5\%) \tabularnewline
        Recruitment & 8 & (1.4\%)\tabularnewline
        Media \& Advertising & 7 & (1.2\%)\tabularnewline
        Construction \& Real Estate & 3 & (0.5\%) \tabularnewline
        None & 41 & (6.9\%)\tabularnewline
        \hline
        Overall & 591 & (100\%)\tabularnewline
		\hline
	\end{tabular}
\end{table}

\begin{table}[!ht]
    \centering
	\caption{Company Size of Industry-Affiliated Respondents}
	\label{table:dems_company_size}
	\begin{tabular}{
	p{0.32\textwidth} 
	>{\raggedleft}p{0.04\textwidth} 
    @{\hspace{1.5\tabcolsep}}
	>{\raggedleft}p{0.05\textwidth} 
	}
 		\# Employees in Company & \multicolumn{2}{r}{\# Respondents}\\ 
        \hline
        \hline
        1-25 & 87 & (15.8\%) \tabularnewline
		26-100 & 87 & (15.8\%) \tabularnewline
        101-1,000 & 143 & (26.0\%) \tabularnewline
        1.001-10,000 & 76 & (13.8\%) \tabularnewline
        \textgreater 10,000 & 157 & (28.5\%) \tabularnewline
        \hline
        Overall & 550 & (100\%) \tabularnewline
		\hline
	\end{tabular}
\end{table}

\noindent
Table~\ref{table:dems_perspective_role} shows the number of participants per role and perspective.
For \textit{hiring}, People Manager / Team Lead was the dominant job role with 115 out of 130 responses.
For the \textit{applicant} perspective, the majority of respondents filled the role of software engineer or architect, with 102 responses being from students of various computer science study programs.
As indicated by Table \ref{table:dems_prof_experience}, the majority of respondents were experienced IT professionals, {\raise.17ex\hbox{$\scriptstyle\sim$}}58\% reported more than 5 years of professional experience.

Concerning the respondents' company domains (see Table~\ref{table:dems_company_domain}), more than 40\% worked for a Software \& IT Services company. 
The remaining portion, however, is spread widely over 14 domains, with Automotive, Manufacturing \& Defense, Research \& Education, and Finance \& Insurance being represented with more than 30 mentions each.
The reported company sizes show a similar diversity (see Table \ref{table:dems_company_size}).
Very small businesses (below 100 employees) are equally represented as large enterprises with more than 10.000 employees, with both groups in the region of 30\%.
As expected, the vast majority of participants stated Germany as their country of residence (529, i.e. {\raise.17ex\hbox{$\scriptstyle\sim$}}90\%).
42 respondents were located in other European countries and 20 on other continents.

\subsection{Hiring Perspective}

\noindent
In this section, we present the results for the \textit{hiring} perspective, which correspond to our research questions 1 and 2.
For RQ1, we were interested in how valuable \textit{hiring} perceives the characteristics \enquote{breadth vs. depth} regarding an applicant's skill set.
We asked for the importance of the following two characteristics while assessing applicants:

\begin{enumerate}
    \item \textbf{broad} knowledge and experience in \textbf{numerous} technologies 
    \item \textbf{deep} knowledge and experience in \textbf{specific} technologies
\end{enumerate}

\begin{figure}[ht]
     \hspace{-0.2cm}
     \includegraphics[keepaspectratio=true, width=0.5\textwidth]{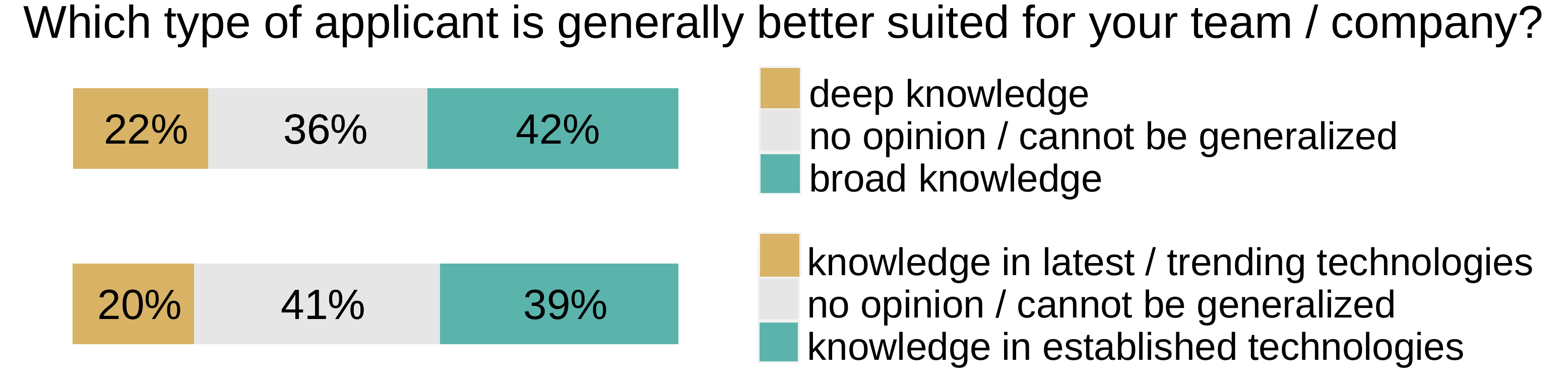}
     \caption{Preferred Skills by Hiring Professionals}
     \label{fig:chart-hiring-prefer}
 \end{figure}

\begin{figure*}[ht]
    \centering
    \includegraphics[width=1.01\textwidth]{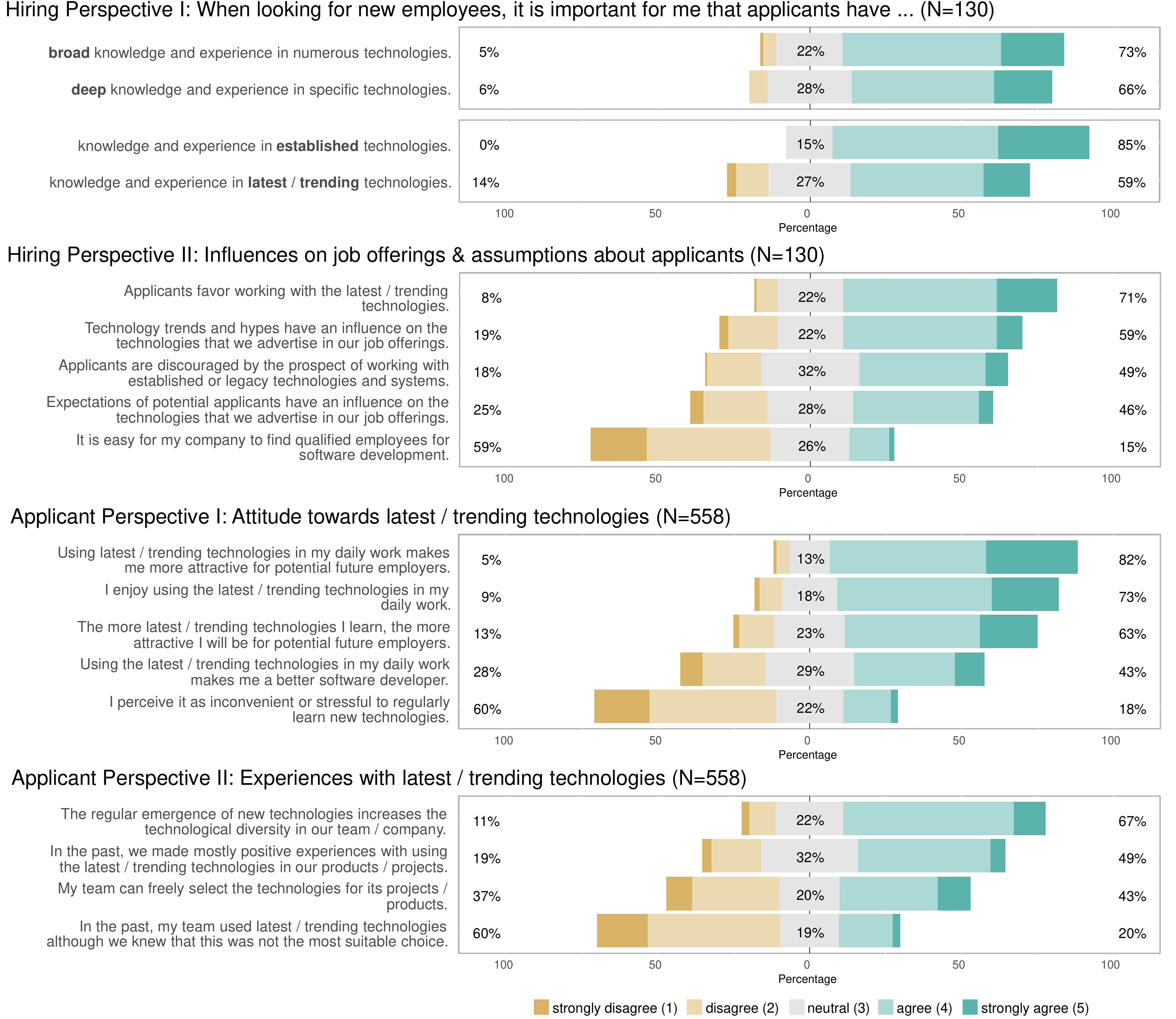}
    \caption{Survey Results by Question Groups for Hiring and Applicant Perspectives}
    \label{fig:chart-likert-results}
\end{figure*}

\noindent 
Respondents assigned fairly equal weights, with \raise.17ex\hbox{$\scriptstyle\sim$}73\% answering \enquote{agree} or \enquote{strongly agree} for \textbf{broad}, while \raise.17ex\hbox{$\scriptstyle\sim$}66\% did the same for \textbf{deep} (see Fig.~\ref{fig:chart-likert-results}).
However, when asked which of the two types would generally be better suited for the their team or company, the results were different.
Forced to decide for one or the other, \raise.17ex\hbox{$\scriptstyle\sim$}42\% now preferred applicants with broad knowledge, while only \raise.17ex\hbox{$\scriptstyle\sim$}22\% chose applicants with deep specific knowledge.
\raise.17ex\hbox{$\scriptstyle\sim$}36\% selected the neutral option, stating that this could not be generalized or that they had no opinion on the matter (see Fig.~\ref{fig:chart-hiring-prefer}).

In the second part of RQ1, we inquired after the perceived importance of experience with latest / trending vs. established technologies in an applicant's skill set.
We again asked for the importance of the following two characteristics during an applicant assessment:

\begin{enumerate}
    \item knowledge and experience in \textbf{latest / trending} technologies
    \item knowledge and experience in \textbf{established} technologies
\end{enumerate}
    
\noindent 
The vast majority (\raise.17ex\hbox{$\scriptstyle\sim$}85\%) confirmed knowledge in \textbf{established} technologies to be important (\enquote{agree} or \enquote{strongly agree}).
For \textbf{latest / trending} technologies, this fraction was considerably lower, but still \raise.17ex\hbox{$\scriptstyle\sim$}59\% (see Fig.~\ref{fig:chart-likert-results}).

\noindent
We again forced the direct comparison by selecting the more suitable characteristic for their team / company: Here only 20\% would prefer the applicant experienced in latest / trending technologies, while \raise.17ex\hbox{$\scriptstyle\sim$}39\% chose experience with established technologies.
However, a similar percentage (\raise.17ex\hbox{$\scriptstyle\sim$}41\%) remained neutral or had no opinion (see Fig.~\ref{fig:chart-hiring-prefer}).

\noindent
For RQ2, we analyzed to which degree \textit{hiring} is influenced by latest / trending technologies when creating job offerings.
A majority of \raise.17ex\hbox{$\scriptstyle\sim$}59\% (\enquote{agree} or \enquote{strongly agree}) confirmed that technology trends and hypes indeed impact the technologies advertised in their job offerings (see Fig.~\ref{fig:chart-likert-results}).
An even larger majority (\raise.17ex\hbox{$\scriptstyle\sim$}71\%) believed that applicants favor working with the latest / trending technologies.
When asked further, if expectations of potential applicants would have an influence on technologies advertised in their job offerings, \raise.17ex\hbox{$\scriptstyle\sim$}46\% confirmed it (\enquote{agree} or \enquote{strongly agree}), while \raise.17ex\hbox{$\scriptstyle\sim$}25\% answered contrarily (\enquote{disagree} or \enquote{strongly disagree}).

In a last question, we asked \textit{hiring} participants if it is easy for their company or customer to find qualified employees for software development
positions.
\raise.17ex\hbox{$\scriptstyle\sim$}59\% answered \enquote{disagree} or \enquote{strongly disagree}, while only \raise.17ex\hbox{$\scriptstyle\sim$}15\% answered \enquote{agree} or \enquote{strongly agree} (see Fig.~\ref{fig:chart-likert-results}), i.e. many of our respondents perceived it as difficult to fill their vacancies.

\subsection{Applicant Perspective}

\noindent
In this section, we present the results for the \textit{applicant} perspective, that correspond to RQ3 and RQ4.
We were interested, to which degree \textit{applicants} are influenced by latest / trending technologies when optimizing their resume and how they perceive hiring attractiveness in this context (RQ3).
Additionally, we wanted to learn about applicants' practical experiences in using latest / trending technologies (RQ4).

When asked, how attractive it is to work with latest / trending technologies, the majority agreed that they enjoy using them in their daily work (\raise.17ex\hbox{$\scriptstyle\sim$}73\%).
Simultaneously, only \raise.17ex\hbox{$\scriptstyle\sim$}18\% perceived it as inconvenient or stressful to regularly learn new technologies (\raise.17ex\hbox{$\scriptstyle\sim$}22\% neutral, \raise.17ex\hbox{$\scriptstyle\sim$}60\% disagree).
However, the belief that using latest / trending technologies in their daily work would make them better software developers was confirmed by only \raise.17ex\hbox{$\scriptstyle\sim$}42\% (see Fig.~\ref{fig:chart-likert-results}).

Complementary to this intrinsic motivation, respondents were also largely convinced that using latest / trending technologies in their daily work makes them more attractive for potential future employers: \raise.17ex\hbox{$\scriptstyle\sim$}82\% answered \enquote{agree} or \enquote{strongly agree}, while only \raise.17ex\hbox{$\scriptstyle\sim$}5\% stated \enquote{disagree} or \enquote{strongly disagree}.
Not quite as unanimous but still confirmed by \raise.17ex\hbox{$\scriptstyle\sim$}63\% was the belief that learning a diverse set of latest / trending technologies would be even more desirable for potential future employers, i.e. \enquote{the more hyped technologies I learn, the more attractive I become} (see Fig.~\ref{fig:chart-likert-results}).

In the last question group, we asked for practical experiences with technology selection and usage.
Nearly half of our participants (\raise.17ex\hbox{$\scriptstyle\sim$}49\%) stated that their past experiences with using latest / trending technologies in products and projects were mostly positive, while \raise.17ex\hbox{$\scriptstyle\sim$}19\% disagreed and \raise.17ex\hbox{$\scriptstyle\sim$}32\% voted neutral (see Fig.~\ref{fig:chart-likert-results}).
Moreover, \raise.17ex\hbox{$\scriptstyle\sim$}20\% reported that they had at some point knowingly selected latest / trending technologies although those had not been ideal for the use case.
We then asked \textit{applicants} to rank the importance of four influencing factors for selecting a technology at the start of a new project.
We calculated the following mean scores from the responses, ranging from 1 (least important) to 4 (most important):

\begin{enumerate}[label=\alph*)]
    \item Functional and non-functional system requirements (3.54)
    \item Skill sets of current employees (3.06)
    \item Latest / trending technologies (1.72)
    \item Skill sets of potential future applicants (1.69)
\end{enumerate}

\noindent
In this ranking, the most reasonable (requirements) and pragmatic (current skills) options were reported as most influential. Latest / trending technologies were ranked third, closely followed by skills of future employees.
Finally, participants were asked if they believed that the regular emergence of new technologies increased technological diversity in their teams or companies: \raise.17ex\hbox{$\scriptstyle\sim$}67\% confirmed this tendency, while only \raise.17ex\hbox{$\scriptstyle\sim$}11\% disagreed and 22\% voted neutral (see Fig.~\ref{fig:chart-likert-results}).

\section{Discussion}

\noindent
In this section, we holistically discuss and interpret the presented survey results.
For both perspectives (hiring and applicant), we review the evidence that supports or refutes the existence of RDD while listing influencing factors based on correlation analysis and regression modeling.

Within the \textit{hiring} perspective, we mainly analyzed two aspects to judge the prevalence of RDD:

\begin{enumerate}[label=\alph*)]
    \item the degree to which technology trends and hypes influence job offerings 
    \item the degree to which expectations of \textit{applicants} influence job offerings.
\end{enumerate}

\noindent
For a), almost 60\% of our participants agreed that trends and hypes influence their job offerings.
For b), the percentage agreement was lower but still half of our participants admitted that applicant expectations influence their job offerings (46\%).
Simultaneously, 71\% believe that applicants favor working with latest technologies and 49\% that applicants are discouraged by legacy technology.
However, our participants also clearly showed a tendency to judge experience with established technologies as overall more important.
In a direct comparison, only 20\% would generally hire the applicant versed in latest / trending technologies.
Moreover, \textit{hiring} generally preferred broad experience with numerous technologies over deep experience with specific technologies, which would be in line with the notion of continuously broadening your skills with e.g. new languages or frameworks.

\begin{figure}[H]
    \centering
    \includegraphics[width=\columnwidth]{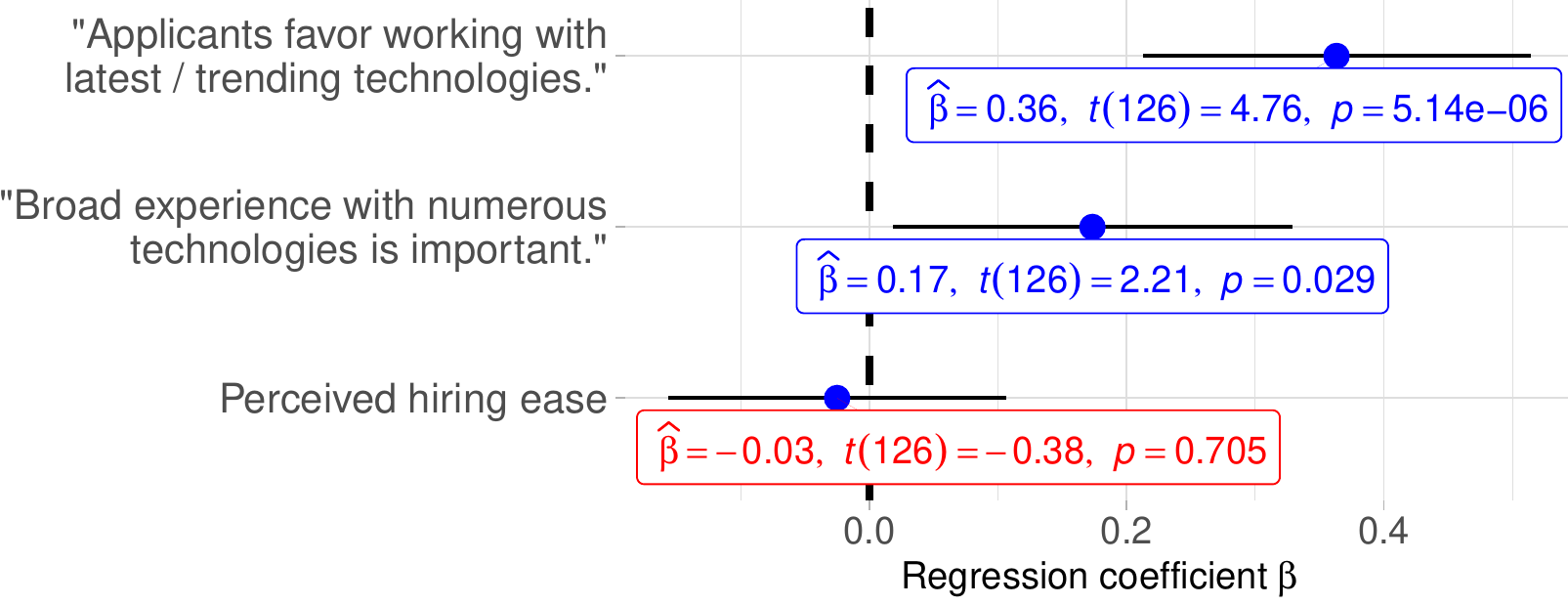}
    \caption{Factors Influencing the RDD Hiring Construct (Linear Regression Model)}
    \label{fig:lm-rdd-hiring}
\end{figure}

\noindent
The combination of the evidence from a) and b) suggests the occurrence of RDD facets in substantial parts of our hiring sample:
even though experience with established technologies is generally preferred, there are strong indications that job offerings are influenced by technology trends and applicant expectations.
To identify influencing factors, we built a linear regression model that tries to predict a combined construct from a) and b), namely the mean value of the influence of trends and hypes on job offerings and the influence of applicant expectations on job offerings.
Predictors for this model were the belief that applicants favor latest technologies and the preference for broad technological experience in applicants (see Fig.~\ref{fig:lm-rdd-hiring}).
For every additional point on one of these Likert scales, the RDD \textit{hiring} construct (also normalized to a 5-point Likert scale) increases by $0.36$ and $0.17$ points respectively.
However, we did not find any additional significant predictors in our \textit{hiring} sample.
No intuitive candidate like the company size, the company perception (from conservative to innovative), or the perceived ease to fill vacancies could be verified.
While the overall model has a p-value of \num{1.553e-05}, it is only able to explain less than 20\% of the variability of the construct (adjusted $R^2=0.159$), i.e. there are other influencing factors we did not capture.

\noindent
For the \textit{applicant} perspective, the occurrence of RDD was mainly judged by two aspects:

\begin{enumerate}[label=\alph*)]
    \item the degree to which \textit{applicants} believe that experience with latest / trending technologies makes them more attractive for employers
    \item the degree to which trends and hypes influence technology selection
\end{enumerate}

\noindent
In contrast to the expressed preference of our \textit{hiring} professionals for experience with established technologies, the vast majority of our \textit{applicant} participants believed to be more hireable via trends and hypes:
82\% agreed that using latest / trending technologies in their daily work makes them more attractive for employers and 63\% even agreed that the more of these technologies they learn the more attractive they become.
This suggests substantial RDD effects that incentivize to keep up-to-date with trends and hypes, especially when considering that only 43\% believe that using latest / trending technologies in their daily work actually makes them better developers.
Considering aspect b), however, the influence of both system requirements ($3.54$) and the skill sets of current employees ($3.06$) were ranked considerably higher than latest / trending technologies ($1.72$).
Simultaneously, just 20\% confirmed to have used hyped technologies in an unsuitable way, although only 49\% stated that they made mostly positive experiences with such technologies.

\begin{figure}[ht]
    \centering
    \includegraphics[width=\columnwidth]{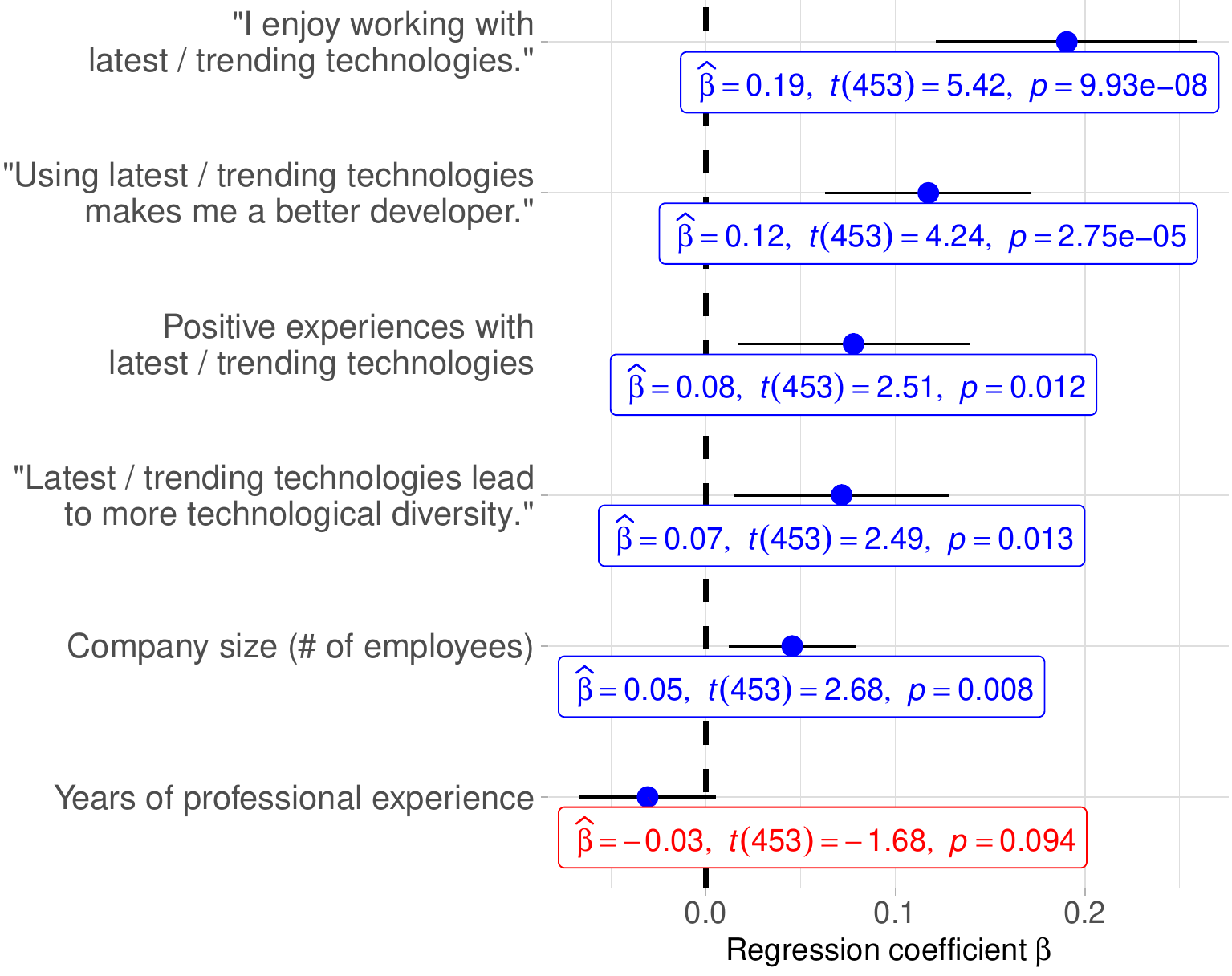}
    \caption{Factors Influencing the RDD Applicant Construct (Linear Regression Model)}
    \label{fig:lm-rdd-applicant}
\end{figure}

\noindent
In combination, a) and b) suggest that RDD facets are similarly prevalent in our \textit{applicant} sample as in our \textit{hiring} sample.
However, the strong evidence from a) and weaker evidence from b) may mean that \textit{applicant} notions of RDD are frequently latent without actually being acted out during professional technology selection.
One reason for this could be that only 43\% of our participants agreed that they can freely select technologies in their projects.
When building a similar linear regression model (see Fig.~\ref{fig:lm-rdd-applicant}) to predict a combined RDD \textit{applicant} construct (the mean value of the degree to which \textit{applicants} believe to become more hireable via using latest technologies in their daily work and the influence of hypes on technology selection), we identified the enjoyment of using latest / trending technologies ($\beta = 0.19$) and the belief to become a better developer by using latest / trending technologies ($\beta = 0.12$) as the top predictors.
Smaller influencing factors were positive experiences with trending technologies, the belief that trends and hypes lead to more technological diversity, and the company size.

The intuitive notion that RDD would be more prevalent with more inexperienced developers could not be fully verified in our sample: while years of professional experience showed some weak correlations with RDD \textit{applicant} facets like the belief that using latest technologies daily increases attractiveness for employers ($r_s = -0.198$, $p = \num{3.8e-06}$), it was not a significant predictor when considering other influences in our linear model ($p = 0.094$).
Likewise, being a student had no significant impact.
The model has a p-value of \num{2.2e-16}, but is only able to explain roughly one third of the variability of the construct (adjusted $R^2=0.311$).

\section{A Theory of Résumé-Driven Development}
\label{sec:final_theory}
\begin{figure*}[ht]
    \centering
    \captionsetup{width=.95\linewidth}
    \includegraphics[width=0.98\textwidth]{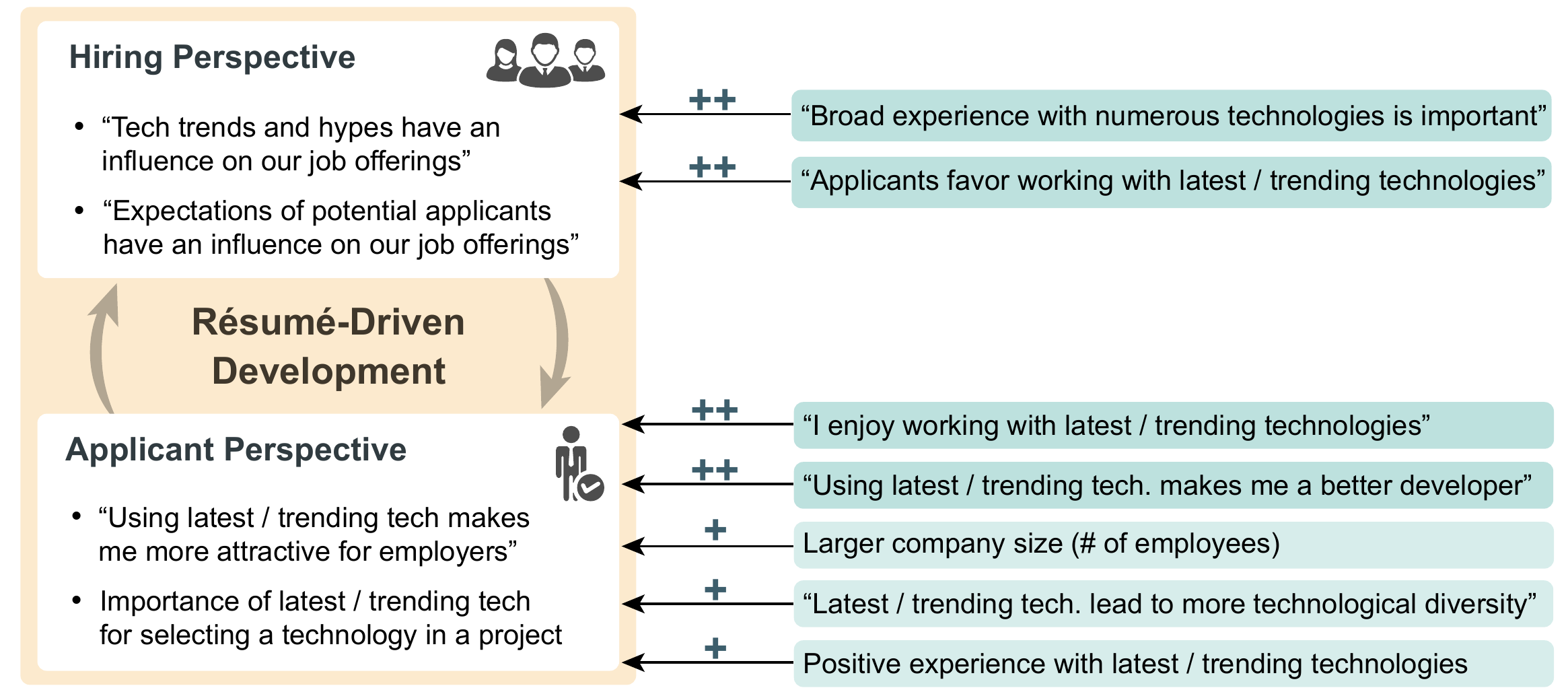}
    \caption{Theory of Résumé-Driven Development. Human resource and software professionals take the perspectives of \textit{hiring} and \textit{applicant} respectively, which are characterized by their facets in RDD. Strengthening predictors are linked to each perspective via + for $\hat{\beta} < 0.1$ and ++ for $\hat{\beta} > 0.1$ (see Fig. \ref{fig:lm-rdd-hiring} and \ref{fig:lm-rdd-applicant}). Icons from~\cite{NounProject2020a}.}
    \label{fig:chart-rdd-theory}
\end{figure*}

\noindent
Our results suggest that RDD facets exist in substantial parts of our population.
Using the framework from Gregor~\cite{Gregor2006}, we propose an analytic theory (Type I) to describe the phenomenon.
We use text and a diagram  for its representation (see Fig.~\ref{fig:chart-rdd-theory}). 
Central constructs are the two perspectives, shaped by their characterizing attributes, and the influencing factors that impact the strength of RDD.
As such, the RDD construct comprises the \textit{hiring} and the \textit{applicant} perspectives.
An influencing factor \textit{strengthens}~(+) the construct of one perspective.
The four aspects with a regression coefficient greater than 0.1 are tagged with \enquote{++}.
The probability that practitioners exert facets of RDD \textit{increases} with seven aspects (see Fig. \ref{fig:chart-rdd-theory}), two for the \textit{hiring} and five for the \textit{applicant} perspective.

The main point expressed by our theory is the technology focus of both parties when creating job advertisements and resumes respectively.
We can observe a tendency by \textit{hiring} to cater to the preferences of \textit{applicants}, who mostly enjoy using new technologies.
However, in reality, such manipulative advertisements are not fully in line with the actual demand of employers. \textit{Applicants}, on the other hand, may react to this technology focus even stronger, promoting them in their resumes in turn. 
RDD may thus develop a momentum of its own.
While it seems intuitive to suspect the shortage of skilled developers as a driver for RDD, we found no related predictors to be significant in our sample (e.g. the perceived hiring ease).
In summary, we define the phenomenon as follows:

\vspace{0.15cm}
\begin{boxC}
\noindent\textbf{Definition.}
\textit{Résumé-Driven Development (RDD) is an \mbox{interaction} between human resource and software professionals in the software development recruiting process.
It is characterized by overemphasizing numerous trending or hyped technologies in both job advertisements and CVs, although experience with these technologies is actually perceived as less valuable on both sides. RDD has the potential to develop a self-sustaining dynamic.
}
\end{boxC}
\vspace{0.1cm}

\noindent
The scope of our theory has been described in Section~\ref{sec:background} and is also subject to our limitations (see Section~\ref{sec:threats}). However, we are confident that our proposed theory at least partially describes a prevalent trend in the software industry as a whole.

\section{Potential Consequences of RDD}

\noindent
Potential consequences of RDD have to be considered in two main areas.
First, the described phenomenon may negatively impact software quality.
Our participants widely agreed that the regular emergence of new technologies increases technological diversity in their companies.
While we did not ask if they perceived this heterogeneity as negative, RDD seems to contribute to the software complexity issue mentioned in the introduction.
A high degree of technological heterogeneity may impact maintainability:
technologies need to be regularly updated, dependencies have to be managed, and knowledge sharing efforts among team members increase with every additional language or framework.
Moreover, introducing new technologies implies a learning curve and may come with maturity issues that impact reliability.
Extensive RDD-based technology selection may therefore lead to complex or even unmaintainable software consisting of technologies which are not suitable for the requirements, which are unfamiliar to current or future employees, or which did not deliver on their promise and were discontinued.

Second, RDD can lead to false expectations and disappointment in the recruiting process on both sides.
In a recent HackerRank survey among 71,281 developers~\cite{HackerRank2020}, the top answer for \textit{\enquote{What turns developers off from employers?}} with an affirmation of more than two-thirds was \textit{\enquote{not enough clarity on role or where I’ll be placed}}.
Similarly, Behroozi at al.~\cite{Behroozi2020} identified a number of suboptimal practices which may sabotage recruitment processes in a study of over 10,000 Glassdoor reviews.
A prevalent theme among them were inadequately communicated criteria by HR.
RDD-based hiring can lead to similar frustrations.
A strong focus on technologies during hiring may also lead to the neglection of other important skills for creating high-quality software products in a team.
This is reinforced by nine \textit{hiring} professionals who provided free-text comments about applicant traits they value much more than experience with trending technology, e.g. soft skills like communication, self-motivation, the willingness to learn, being a cultural fit for the team, or an understanding of the fundamental principles behind technologies.
Since high employee turnover can be especially harmful in knowledge-intensive fields like software development, potential consequences of RDD hiring should not be underestimated.

\section{Threats to Validity}
\label{sec:threats}

\noindent
Limitations of our study have to be mentioned in several areas.
One threat to construct validity is the absence of similar works or questionnaires we could build on.
While we carefully reviewed related literature, used proven Likert scales, and followed established guidelines, there is still a chance that some of our questions did not suitably capture what we actually wanted to measure.

Several precautions were taken to ensure internal validity.
To mitigate consistency and representation issues, we piloted and revised the questionnaire several times and also consulted other researchers.
Some respondents may still have given inaccurate or untrue answers, either deliberately, due to misunderstandings, or due to cognitive bias.
As mitigation strategies, we defined important terms at the start of the questionnaire.
To create a common understanding of the involved concepts, our description of \enquote{latest / trending technologies} also included \enquote{hyped technologies}, which some may interpret as a derogatory term.
Participation was also strictly voluntary and we assured anonymity for further processing and publishing of the collected data.
The English questionnaire could have influenced non-native speakers, but we expect this to be minimal in the software engineering field.
Lastly, we did not use the term RDD to avoid bias in participants.

A threat to conclusion validity could arise from the exploratory nature of our survey. As such, we did not have formal hypotheses, e.g. for testing if RDD exists, but relied on our interpretation of distributions.
To minimize researcher bias, we discussed all important results between the first three authors until consensus was reached.
A confounding factor for answers to the \textit{applicant} perspective could be prescribed technology choices by employers, as reported by 43\% of our participants.
We may have missed other such factors for the phenomenon, which is also supported by the low degree of variance our regression models are able to explain.
Moreover, while using linear regression instead of simple correlation analysis improved our understanding of relationships in our data, we still cannot make statements about causality.

For scientific SE surveys, we had a high number of 558 responses for the \textit{applicant} and 130 for the \textit{hiring} perspective, with diversity in work experience, company size, and domain.
A threat to external validity, i.e. generalizability, may be that the majority of participants were located in Germany (\raise.17ex\hbox{$\scriptstyle\sim$}90\%).
Regional and cultural factors could influence the phenomenon and results could partially differ in a sample dominated by participants from, e.g., the US.
Furthermore, our \textit{applicant} sample contained 102 students (18\%). While we think that the views of students are also relevant for the \textit{applicant} perspective, they may differ from the opinions of professionals. 
Nonetheless, we still believe the fundamentals of our theory to be applicable to a broad range of software engineering contexts.

\noindent
Finally, our survey was conducted during the lock-downs related to COVID-19 in 2020, confronting human society with significantly changed living and working conditions.
We still believe that responses were not seriously affected, since related opinions should have formed over a longer timeframe.
We also assume that the long-term economic impact will be limited and thus not drastically change the characteristics of the investigated phenomenon.

\section{Conclusion}

\noindent
In this study, we empirically characterized and defined the phenomenon Résumé-Driven Development (RDD) by interpreting results of a survey with 591 hiring and software professionals.
The survey confirmed the occurrence of RDD facets in substantial parts of our sample. 
\textit{Hiring} professionals valued a broad technical skill set in \textit{applicants}.
The majority of them agreed that \textit{applicants} favor working with latest / trending technologies and that those technologies influence their job offerings.
Simultaneously, they preferred experience with established technologies in \textit{applicants}.
The majority of our \textit{applicants}, on the other hand, believed that using latest technologies in their daily work makes them more attractive for employers and still more than half believed that the more of these technologies they learn the more attractive they become.
However, less software professionals reported negative experiences with or unsuitable selection of such technologies.

Based on the results, we proposed a descriptive theory to scope and partially explain RDD.
The theory contains a definition, the characterization of the two RDD perspectives, and influencing factors that strengthen or weaken RDD.
Potential consequences of RDD are mainly decreased software quality 
and increased employee turnover due to false expectations on both sides.

Future research needs to confirm or reject this theory, i.e. more evidence is needed to analyze the phenomenon.
Replicating this survey, especially in different geographical and cultural settings, may provide valuable insights to adapt and extend the theory, e.g. the influences and consequences.
Furthermore, the rich results of qualitative interviews may shed more light on practitioners' thought processes and rationales.
Overall, we think that an understanding of RDD, its prevalence, and its consequences are important in a time where new technologies, frameworks, and languages are released more and more frequently.

\section{Data Availability}
\noindent
We openly share the questionnaire, anonymized raw data, and R script for the analysis.\footnote{https://doi.org/10.6084/m9.figshare.c.5134226.v1}

\section*{Acknowledgment}
\noindent
The authors would like to thank Maximilian Jager, University of Mannheim, for his consultation on the used statistical methods and Daniel Graziotin, University of Stuttgart, for his feedback on the questionnaire and paper.

\bibliographystyle{IEEEtran}
\bibliography{references.bib}

\end{document}